# Spin-order-induced multiferroicity in LiCuFe$_2$(VO$_4$)$_3$ and disorder effects in NaCuFe$_2$(VO$_4$)$_3$

A.V. Koshelev,[1,2] K.V. Zakharov,[1] L.V. Shvanskaya,[1,3] A.A. Shakin,[3] D.A. Chareev,[2,4] S. Kamusella,[5] H.-H. Klauss,[5] K. Molla,[6] B. Rahaman,[6] T. Saha-Dasgupta,[7] A.P. Pyatakov,[1] O.S. Volkova,[1,2,8] and A.N. Vasiliev[1,3,8]

[1]Lomonosov Moscow State University, 119991 Moscow, Russia
[2]Ural Federal University, 620002 Ekaterinburg, Russia
[3]National University of Science and Technology "MISiS", 119049 Moscow, Russia
[4]Institute of Experimental Mineralogy, RAS, 142432 Chernogolovka, Russia
[5]Dresden Technical University, 01062 Dresden, Germany
[6]Aliah University, Kolkata, India
[7]S.N. Bose National Centre for Basic Sciences, Kolkata, India
[8]National Research South Ural State University, 454080 Chelyabinsk, Russia

Mixed spin chain compounds, ACuFe$_2$(VO$_4$)$_3$ (A= Li,Na), reach magnetically ordered state at $T_N$ ~ 11 K (Li) or ~ 9 K (Na) and experience further transformation of magnetic order at T* ~ 7 K (Li) or ~ 5 K (Na), evidenced in magnetic susceptibility $\chi$ and specific heat $C_p$ measurements. While no anomaly has been detected in dielectric property of NaCuFe$_2$(VO$_4$)$_3$, the step-like feature precedes a sharp peak in permittivity $\varepsilon$ at $T_N$ in LiCuFe$_2$(VO$_4$)$_3$. These data suggest the spin-order-induced ferroelectricity in Li compound and no such thing in Na compound. On the contrary, the Mössbauer spectroscopy study suggests similarly wide distribution of hyperfine field in between T* and $T_N$ for both the compounds. The first principles calculations also provide similar values for magnetic exchange interaction parameters in both compounds. These observations lead us to conclude on the crucial role of alkali metals mobility within the channels of the crystal structure needed to be considered in explaining the improper multiferroicity in one compound and its absence in other.

**Introduction**

The coexistence of ferroelectricity and magnetism being considered for a long time mutually exclusive phenomena[1] represent now a flourishing field of magnetoelectric multiferroicity.[2-4] Quite a number of sophisticated physical mechanisms may justify the coexistence (and coupling) of these order parameters in a single homogeneous material.[5-8] Phenomenologically, multiferroics can be divided in two classes: type-I multiferroics with ferroelectric Curie point higher than the magnetic ordering temperature, and type-II (or improper) multiferroics with magnetically induced electric polarization appearing below magnetic Curie or Neel temperatures.[9] In particular, the magnetoelectric coupling is an intrinsic feature of the improper multiferroics where the ferroelectric polarization arises from some spiral magnetic ordering.[10-13] The magnetic spirals are the result of competition of the nearest neighbor and the higher order superexchange interactions. Thus, choosing the compound with the appropriate bond angles in three-ion clusters (transition-metal ion – ligand – transition metal ion) allows one to fine-tune the balance between the exchange interactions simply by varying the temperature. Instructive in this sense is the case of cupric oxide, CuO, which reaches collinear long-range magnetic order through succession of two phase transitions at $T_{N1}$ = 213 K and $T_{N2}$ = 230 K with incommensurate spiral magnetic structure in between these temperatures.[14] The phase transition at $T_{N2}$ is marked by sharp peak in dielectric permittivity reflecting the formation of ferroelectric phase.

Here, we report on observation of spin-order-induced ferroelectricity in LiCuFe$_2$(VO$_4$)$_3$ fully absent in its mineral counterpart NaCuFe$_2$(VO$_4$)$_3$ (howardevansite). In many respects, the howardevansites differ from any multiferroic material studied so far. Their magnetic subsystem is represented by two types of 3d-transition metal, Cu and Fe, with third 3d-transition metal, V, being magnetically silent. The alkali metals, Li or Na, are subject to hopping within the tunnels of crystal structure. Belonging to the space group $P\bar{1}$, this compound may serve as an example of type-II multiferroic medium with the lowest possible symmetry of paraelectric and multiferroic phases: the only symmetry element is the inversion center that is broken at Neel temperature.

$ACuFe_2(VO_4)_3$ has a triclinic primitive cell with two formula units per unit cell, consisting of two inequivalent Fe atoms Fe1 and Fe2, which are situated at the centre of a distorted octahedra constructed by six O-atoms and one Cu atom which is at the centre of a distorted trigonal bipyramid constructed by five O-atoms. In $ACuFe_2(VO_4)_3$, two distorted $Fe1O_6$ octahedra share a common edge and form $Fe1O_6$ dimers. The dimers of $Fe2O_6$ octahedra form similarly. These dimers are connected by edge sharing via distorted $CuO_5$ trigonal bipyramids and form Fe1-Fe1-Cu-Fe2-Fe2 mixed spin chain. The chains are connected by $VO_4$ tetrahedra forming a three dimensional network. The Li or Na ions occupy the channels of this structure. One of two alkali ions position is half-filled.[15] The crystal structure of $ACuFe_2(VO_4)_3$ in polyhedral representation is shown in Fig. 1. The synthesis and physical properties of $LiCuFe_2(VO_4)_3$ were reported in Ref. 16. It was concluded that this compound is a kind of quasi-one-dimensional antiferromagnet. No evidence for spin-order-induced multiferroicity was presented, however.

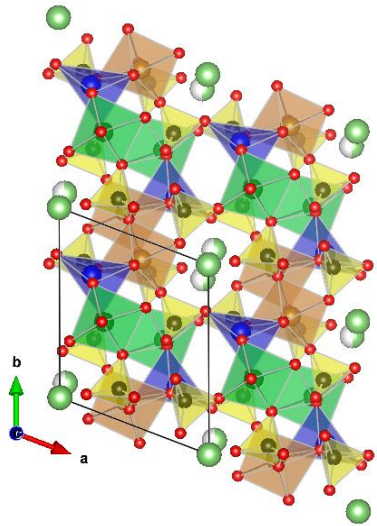

Fig. 1. Crystal structure of $ACuFe_2(VO_4)_3$ (A = Li, Na) in *ab* plane. Two different alkali ion positions are shown by isolated spheres.

**Experiment and calculation details**

The powder samples of $ACuFe_2(VO_4)_3$ (A = Li, Na) were prepared by solid-state reaction of $Na_2CO_3$ or $Li_2CO_3$, $Fe_2O_3$ (preliminary burned under vacuum), CuO and $V_2O_5$, following the procedure described in Ref. 17. The reagents were taken in stoichiometric ratio, homogenized in an agate mortar, placed into the alumina crucibles and burned in two steps in air at 350 and 600 C (Na compound) and 600 and 640 C (Li compound) for 72 hours with intermediate grinding. The crystalline quality and phase purity of the samples obtained were confirmed by the powder XRD method using ADP Diffractometer (Co $K_\alpha$ radiation, $\lambda = 1.789010$ Å).

Thermodynamic properties of $ACuFe_2(VO_4)_3$, i.e. specific heat $C_p$, *ac*- and *dc*-susceptibility χ were measured in the range 2 – 300 K using relevant options of "Quantum Design" Physical Property Measurement System PPMS – 9T. Dielectric permittivity ε was measured in the frequency range f = 10 ÷ $10^3$ Hz by means of a capacitance bridge Andeen-Hagerling 2700A. No frequency dependence of dielectric property in both compounds was detected.

Mössbauer spectroscopy was done with a standard Wissel setup, using a $^{57}$Co source with initial activity of 1.67 GBq and a source line width of 0.107 mm/s. The drive was run in sinusoidal mode minimizing the velocity error. The measurements were carried out in Oxford continuous flow cryostat in under-pressure mode. The typical standard deviation in temperature is 0.01 K, a maximum deviation of 0.03 K is guaranteed. The spectra were analyzed using Moessfit algorithm by means of transmission integral fits.[18]

In order to gain microscopic understanding of ACuFe$_2$(VO$_4$)$_3$ (A=Li, Na), we carried out first principles density functional theory (DFT) calculations[19] within the generalized gradient approximation (GGA) for the exchange-correlation functional.[20] Calculations have been carried out using plane wave basis as implemented within Vienna Ab initio Simulation Package (VASP)[21] as well as muffin-tin orbital (MTO) based *N*-th order MTO (NMTO)[22] and linear MTO (LMTO)[23] calculations implemented in Stuttgart code. The effect of Coulomb correlation beyond GGA at the transition metal sites (Fe and Cu) has been handled within supplemented +U correction of GGA+U.[24] The consistency of obtained results between the muffin-tin orbital basis set calculations and plane wave basis set calculations has been checked in terms of ground state electronic structures.

**Thermodynamic properties**

The temperature dependences of specific heat $C_p/T$, *dc*-susceptibility $\chi_{dc}$, *ac*-susceptibility $\chi_{ac}$ and dielectric permittivity $\varepsilon$ in ACuFe$_2$(VO$_4$)$_3$ (A = Li, Na) are shown in Fig. 2. Despite obvious similarities in behavior of these compounds, there are also features, which allow distinguishing the Li and Na species. Basically, both these compounds experience long range antiferromagnetic order at low temperatures accompanied by distinct anomalies in specific heat and magnetic susceptibility. The parameters of the magnetic subsystem in ACuFe$_2$(VO$_4$)$_3$ estimated in the range 150 – 300 K are collected in Table 1. The large values of the so-called frustration ratio, $|\Theta|/T_N \sim 10$, where $\Theta$ is the Weiss temperature and $T_N$ is the Neel temperature, point to reduced dimensionality of the magnetic subsystem in howardevansites. The values of the Curie constant $C$ are consistent with notion of divalent copper Cu$^{2+}$ (S = 1/2) and pentavalent iron Fe$^{3+}$ (S = 5/2). The negative and comparable values of the temperature-independent term $\chi_0$ are due to the summation of diamagnetic Pascal constants of individual ions in the structure of ACuFe$_2$(VO$_4$)$_3$.[25]

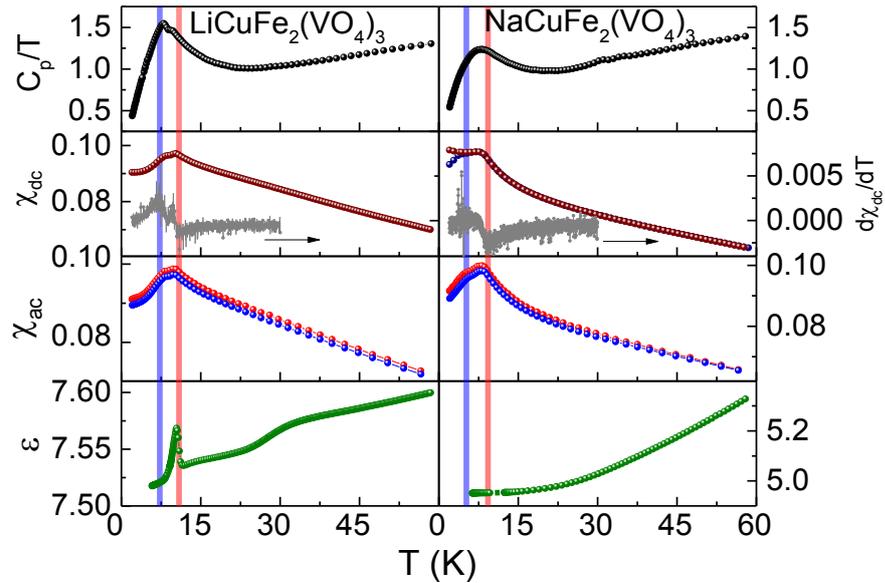

Fig. 2. The temperature dependences of specific heat $C_p/T$, *dc*-susceptibility $\chi_{dc}$, *ac*-susceptibility $\chi_{ac}$, and permittivity $\varepsilon$ in ACuFe$_2$(VO$_4$)$_3$ (A = Li, Na).

At low temperatures, $C_p/T$ vs $T$ curves evidence two separated peaks in close vicinity to each other in LiCuFe$_2$(VO$_4$)$_3$ and broad hump in NaCuFe$_2$(VO$_4$)$_3$. The two peak structure of magnetic singularity is further confirmed in measurements of dc- and ac-susceptibility. In this property, it is seen in both compounds. The upper peak's temperature $T_N$ = 10.9 K in Li compound and $T_N$ = 9.3 K in Na compound can be defined as Neel temperature. The lower peak's temperature $T^*$ = 7.2(5) K in Li compound and $T^*$ = 5.2 K in Na compound marks, presumably, incommensurate-commensurate phase transition. The similarity of the magnetic response in both compounds follows from $d\chi_{dc}/dT$ vs $T$ curves shown in Fig. 2. While no difference between zero-field-cooled and field-cooled $\chi_{dc}(T)$ curves was detected in LiCuFe$_2$(VO$_4$)$_3$, these curves spread at $T < T^*$ in NaCuFe$_2$(VO$_4$)$_3$. The $\chi_{ac}(T)$ curves taken at the frequencies $10^2$ Hz (red symbols), $10^3$ Hz (not shown) and $10^4$ Hz (blue symbols) do not evidence any qualitative difference in behavior of Li and Na compounds. Overall, both magnetic susceptibility and specific heat hint on some disorder in NaCuFe$_2$(VO$_4$)$_3$ to be compared to LiCuFe$_2$(VO$_4$)$_3$.

What differs these two species is the dielectric permittivity $\varepsilon$. At lowering temperature, it smoothly goes down in Na compound and exhibit (i) step-like feature at about 30 K and (ii) sharp asymmetric peak at $T_N$ in Li compound. The absolute value of dielectric permittivity $\varepsilon$ in LiCuFe$_2$(VO$_4$)$_3$ is one and a half larger than $\varepsilon$ in NaCuFe$_2$(VO$_4$)$_3$. This fact can be associated with the enhanced mobility of Li ions within the channels of the crystal structure compared to mobility of Na ions. The step-like feature in dielectric response of Li compound can be ascribed tentatively to the freezing or ordering of the lithium ions subsystem. The peak in $\varepsilon(T)$ curve at $T_N$ is reminiscent of that in CuO, where it was associated with formation of the incommensurate magnetic structure at $T_N$ = 230 K. Note, no anomaly was detected in dielectric property at commensurate-incommensurate phase transition at $T^*$ = 213 K.[14] The absence of sharp anomalies in dielectric permittivity $\varepsilon$ of NaCuFe$_2$(VO$_4$)$_3$ can be due to residual disorder in the less mobile sodium ions subsystem.

**Mössbauer spectroscopy**

The Mössbauer spectra of ACuFe$_2$(VO$_4$)$_3$ (A=Li,Na) taken at various temperatures both in paramagnetic and magnetically ordered states are shown in Fig. 3. A two site model was applied to reproduce the data. Both sites are based on the static Hamiltonian

$$H_s = \frac{eQ_{zz}V_{zz}}{4I(2I-1)}\left[(3I_z^2 - I^2) + \frac{\eta}{2}(I_+^2 + I_-^2)\right] - g_I\mu_N B\left(\frac{I_+ e^{-i\Phi} + I_- e^{+i\Phi}}{2}\sin\Theta + I_z\cos\Theta\right) \quad (1)$$

sharing the same isomer shift. For the electric field gradient (EFG) axial symmetry was assumed, $\eta = 0$. The angle between EFG z-axis and Bhyp is 90° in this model. The principal component of EFG was assumed temperature independent and the same for both sites $V_{zz}$ = –39 V/Å$^2$ = 0.65 mm/s in LiCuFe$_2$(VO$_4$)$_3$ and $V_{zz}$ = –43.5 V/Å$^2$ = 0.725 mm/s in NaCuFe$_2$(VO$_4$)$_3$.

The spectra require a real distribution of static magnetic hyperfine field $B_{hyp}$ because already at base temperature (2 K) they show strongly asymmetric line shape. The analysis was done in a purely static picture using the following theory

$$m \in [-1,1]; \ \sigma \in [0, B_0/(1/2 + m/6)]; \ B_{min} = B_0 - (1/2 + m/6)\sigma; \ B_{max} = B_0 + (1/2 - m/6)\sigma;$$

$$x = \frac{B_{hyp} - B_{min}}{B_{max} - B_{min}} \Rightarrow x \in [0,1]; \ y_0 = (1-m); \ \rho(x) = 2mx + y_0;$$

$$<B_{hyp}> = B_{hyp}(<x>) = B_{hyp}\left(\int_0^1 x\rho(x)dx\right) = B_{hyp}(x = 1/2 + m/6) = B_0 \quad (2)$$

This distribution is linear in $B_{hyp}$ and can change its weight from lower fields ($m = -1$) to higher fields ($m = +1$). Thus, the distribution is parametrized by the mean value $B_0$, the distribution width $\sigma$ and the skewness $m$. This model can roughly satisfy all Mössbauer spectra. To describe quantitatively the Mössbauer spectra in both compounds we had to introduce an impurity sextet, which is still present at room temperature. Its fraction weight is about 5% in LiCuFe$_2$(VO$_4$)$_3$ and about 7% in NaCuFe$_2$(VO$_4$)$_3$.

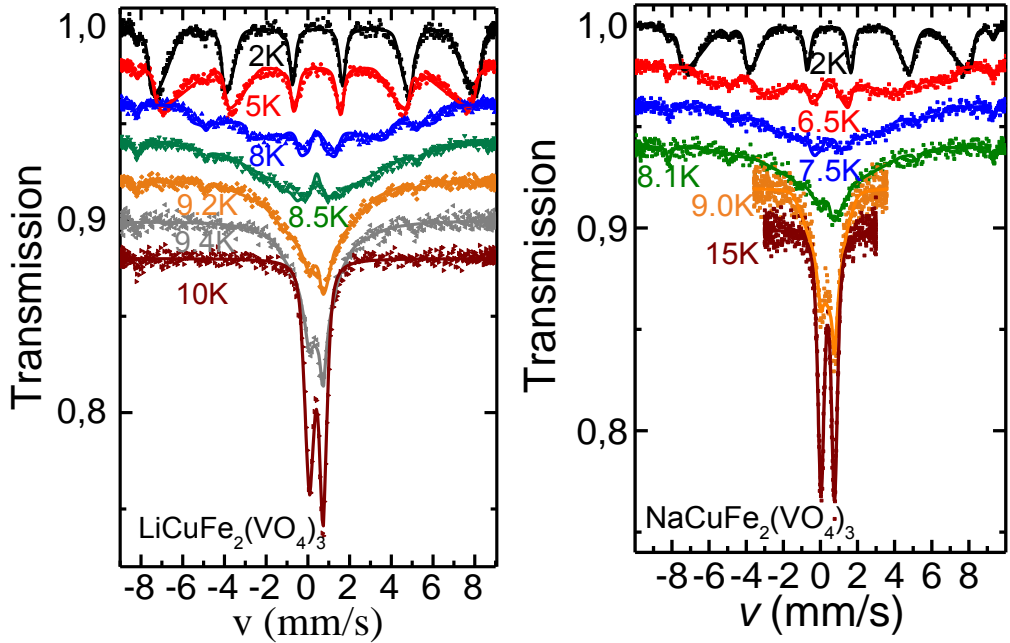

Fig. 3. The temperature evolution of the Mössbauer spectra in LiCuFe$_2$(VO$_4$)$_3$ (left panel) and NaCuFe$_2$(VO$_4$)$_3$ (right panel). The solid lines are the fits with static field distribution, as given by Eq. 2.

For both ACuFe$_2$(VO$_4$)$_3$ (A=Li,Na) the data indicate substantional changes in the field distribution by the change of the skewness m from −1 to +1 at about 8.2 K in Li compound and at about 7.2 K in Na compound. The low temperature isomer shift with respect to room temperature iron δ(T→0) ≈ 0.5 mm/s is the same in both howardevansites. The model roughly fits the data and provides a temperature dependence of the parameters $B_0$ and σ, as shown in Fig. 4 (left panel). The magnetic transition is about 1 K lower in NaCuFe$_2$(VO$_4$)$_3$ as compared to LiCuFe$_2$(VO$_4$)$_3$, but there are no significant differences concerning the absolute value of the saturation hyperfine field $B_0$(T→0) or the width σ.

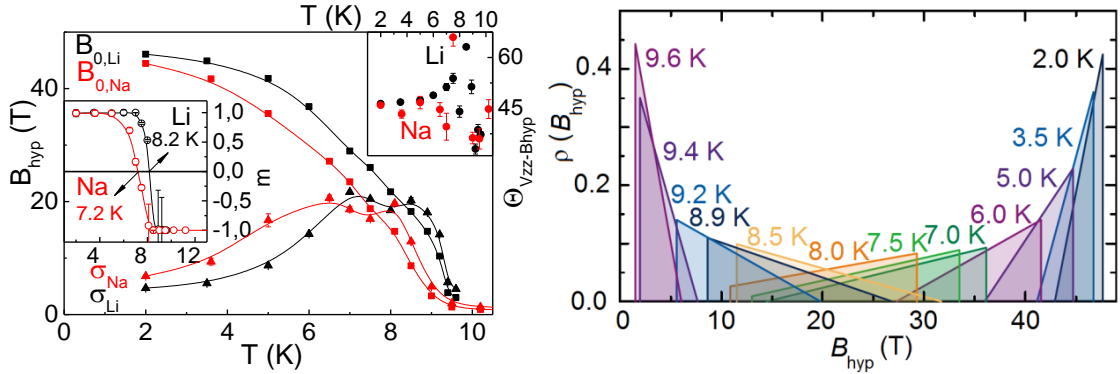

Fig. 4. Magnetic hyperfine field distribution in ACuFe$_2$(VO$_4$)$_3$ modelled with Eq. 2 (left panel). Distribution of hyperfine field at $^{57}$Fe nuclei in LiCuFe$_2$(VO$_4$)$_3$ (right panel).

The non-monotonous behavior of the width parameter σ is evident in the range between $T^*$ and $T_N$ in both compounds. A plot sketching the modelled field distribution in LiCuFe$_2$(VO$_4$)$_3$ at various temperatures in the range between $T^*$ and $T_N$ is shown in Fig. 4 (right panel). This kind of hyperfine field distribution at $^{57}$Fe is consistent with commensurate-incommensurate scenario suggested by ε(T) data for Li compound.

**First principles calculations**

Non-spin-polarized self-consistent calculations have been carried out for $ACuFe_2(VO_4)_3$ (A=Li,Na) compounds. The states close to Fermi level are dominantly of the Fe-d and Cu-d characters with admixture of the O-p character due to finite Fe−O and Cu-O covalency. Thus, we conclude that Fe-d and Cu-d states at Fermi level are primarily responsible for the electronic and magnetic behavior of these compounds. The corresponding spin-polarized density of states, obtained in a self-consistent spin polarized DFT calculation, projected onto Fe-d, Cu-d, O-p, V-d, and A-s states indicates that the Fe-d states are completely filled in the majority spin channel and completely empty in the minority channel, suggesting the nominal $Fe^{+3}$ or $d^5$ valence state of Fe while the Cu-d states are completely filled in the majority and minority spin-channels except the minority channel of Cu-$d_{x2-y2}$ suggesting the nominal $Cu^{+2}$ or $d^9$ valence of Cu. The O-p state is found to be mostly occupied suggesting the nominal $O^{-2}$ valence states. The O-p state shows finite, non-zero hybridization with Fe-d and Cu-d states close to Fermi energy, which contributes to the super-exchange paths of magnetic interaction. The oxidation state of A and V ions are +1 and +5 respectively. The calculated magnetic moment at Fe and Cu sites are found to be 4.54 $\mu_B$ and 0.77 $\mu_B$ with rest of the moment sitting at neighboring O site, with a total magnetic moment of 11 $\mu_B$ per formula unit.

**Magnetic Interactions:**

In order to estimate the various Cu-Cu, Fe-Fe, Cu-Fe magnetic exchange interactions as well as crystal field splitting of different d-orbitals of Fe and Cu present in these compounds, $N^{th}$ order Muffin Tin Orbital (NMTO) based downfolding technique was applied to construct Fe-d and Cu-$d_{x2-y2}$ only Wannier functions by downfolding all the degrees of freedom associated with O, V, A and the Cu and keeping active only the Fe-d and Cu-$d_{x2-y2}$ degrees of freedoms. This procedure provides renormalization of Fe-d and Cu-$d_{x2-y2}$ orbitals due to hybridization from O-p, V-d and A-s and the other Cu-d orbitals. The effective Fe-Fe, Fe- Cu $d_{x2-y2}$ and Cu $d_{x2-y2}$ - Cu $d_{x2-y2}$ hopping interactions were obtained from the real space representation of the Hamiltonian in the effective Fe-d and Cu-$d_{x2-y2}$ Wannier function basis. The strengths of hopping interactions provide us the information on dominant Fe-Fe, Fe-Cu and Cu-Cu magnetic interactions. In principle, the magnetic interactions can be calculated from the knowledge of hopping interactions and crystal field splitting, together with a choice of Hubbard U parameter in the superexchange formula. This procedure, however, only accounts for the antiferromagnetic contributions. Thus, to obtain a more reliable estimate of magnetic exchanges, we performed total energy calculation of different spin configurations in GGA+U scheme and extracted the dominant magnetic exchanges by mapping the DFT energies to that of the Heisenberg model. The GGA+U calculations were performed for a choice of U = 8 eV. The spin chain formed by Fe1, Fe2 and Cu spins and the paths for dominant magnetic interactions for $ACuFe_2(VO_4)_3$ (A=Li, Na) compounds are shown in Fig. 5. The calculated values of magnetic exchanges are listed in Table 2.

Focusing on the calculated magnetic interactions, as listed in Table 2, the first thing we notice is that inter-chain interactions are order of magnitude smaller compared to dominant nearest-neighbor intra-chain interactions, suggesting weakly coupled spin chain model of the compounds. Secondly, we find that the quantitative values of the magnetic interactions to be rather similar between the two compounds, suggestive of similar magnetic properties of the two compounds under static condition. Note our calculation does not take into account the possible mobility of Li+ or Na+ ions within the channel, which can modulate the magnetic interactions. We find the dominant magnetic interactions to be of antiferromagnetic nature, though some of the ferromagnetic interactions are also present which are weaker in strength. This may give rise to complex magnetic ground state.

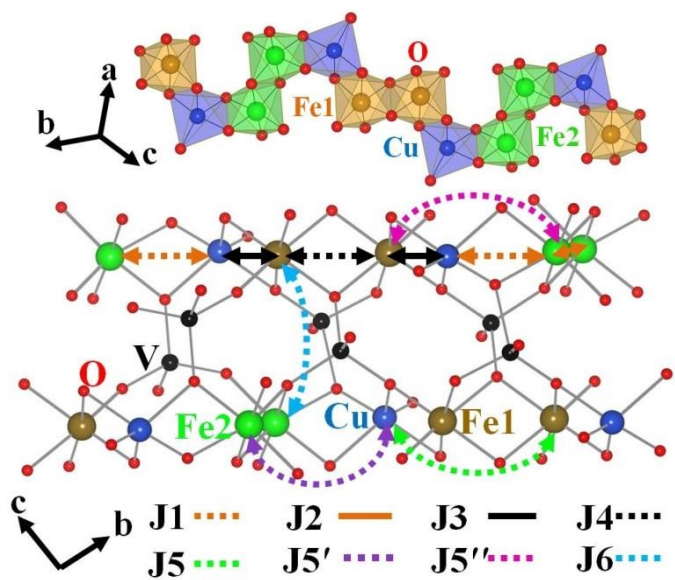

Fig. 5. Mixed spin chain (upper panel) and interaction pathways for dominant magnetic exchange interactions J1−J6 in $ACuFe_2(VO_4)_3$ (A=Li,Na) compounds (lower panel).

**Concluding remarks**

The magnetic subsystem in $ACuFe_2(VO_4)_3$ (A=Li,Na) howardevansites is quite peculiar being constituted by weakly coupled chains of mixed spins, S – S – s – S – S, where small s = 1/2 belongs to $Cu^{2+}$ ions and large S = 5/2 belongs to $Fe^{3+}$ ions. Within chains these spins are coupled by multiple antiferromagnetic (strong) and ferromagnetic (weak) exchange interactions which opens a way to formation of non-collinear structures, either commensurate or incommensurate. The latter ones break the spatial inversion symmetry of the compound. Sharp peak in dielectric permittivity ε suggests the phase transition to multiferroic state accompanying the formation of incommensurate magnetic structure at $T_N$ which is consistent with observed hyperfine field distribution in between $T^*$ and $T_N$. Two phase transitions of magnetic origin seen in specific heat $C_p$ and magnetic susceptibility χ support the scenario of incommensurate – commensurate transformation at $T^*$. The whole of available data points to improper multiferroicity in $ACuFe_2(VO_4)_3$ (in Li species, at least). The hopping of alkali ions along the [001] channels of howardevansite structure was established in electrical conductivity measurements at elevated temperatures.[26,27] These data clearly evidence high mobility of Li ions as compared to Na ions. The presence of mobile subsystem of alkali metals, A = Li or Na, within channels of howardevansite structure brings new colors to the palette of magnetoelectric phenomena. It seems that residual disorder of less mobile Na ions as compared to Li ions may destroy the fine balance of multiple exchange interactions and wipe out the magnetically induced ferroelectricity.

**Acknowledgements**

We acknowledge support of Russian Foundation for Basic research through joint Russia – India project 17-52-45014. This work has been supported also by the Russian Ministry of Education and Science of the Russian Federation through NUST «MISiS» grant K2-2016-066 and by the Act 211 of the Government of Russia, contracts 02.A03.21.0004, 02.A03.21.0006 and 02.A03.21.0011.

Table 1. Parameters of magnetic subsystem in $ACuFe_2(VO_4)_3$ (A = Li, Na).

| Parameter | $LiCuFe_2(VO_4)_3$ | $NaCuFe_2(VO_4)_3$ |
|---|---|---|
| $\chi_0$, emu/mol | $-7\times10^{-4}$ | $-8\times10^{-4}$ |
| $C$, emuK/mol | 9.83 | 9.29 |
| $\Theta$, K | $-80$ | $-86$ |
| $T_N$, K | 10.9 | 9.3 |
| $T^*$, K | 7.2(5) | 5.2 |
| $|\Theta|/T_N$ | 7 | 9 |

Table 2. Magnetic exchange interaction parameters in $ACuFe_2(VO_4)_3$ (A = Li, Na).

| Interaction | Path | $LiCuFe_2(VO_4)_3$ | | | $NaCuFe_2(VO_4)_3$ | | |
|---|---|---|---|---|---|---|---|
| | | Bond angles | Bond distance, Å | Value, meV | Bond angle | Bond-distance, Å | Value meV |
| Nearest-neighbor intrachain | J1 Cu-O-Fe2 | 99.9° & 99.2° | 3.074 | -0.11 | 100.8° & 98.8° | 3.067 | -0.16 |
| | J2 Fe2-O-Fe2 | 104.2° & 104.9° | 3.112 | 1.02 | 106.0° | 3.185 | 1.24 |
| | J3 Cu-O-Fe1 | 98.4° & 103.7° | 3.121 | 1.08 | 102.7° & 100.2° | 3.145 | 1.07 |
| | J4 Fe1-O-Fe1 | 103.5° & 104.4° | 3.156 | 1.38 | 103.0° | 3.096 | 1.33 |
| Next-nearest-neighbor intrachain | J5 Cu-O-Fe1-O-F1 | | 5.122 | -0.09 | | 5.123 | -0.11 |
| | J5′ Cu-O-Fe2-O-F2 | | 5.150 | -0.01 | | 5.176 | -0.01 |
| | J5″ Fe1-O-Cu-O-Fe2 | | 5.114 | 0.05 | | 5.151 | 0.06 |
| Interchain | J6 Fe1-O-V-O-Fe2 | | 4.753 | 0.11 | | 4.760 | 0.13 |